\documentclass[aps,prl,superscriptaddress,amsmath,amssymb,showpacs]{revtex4}
\usepackage{amsmath}
\usepackage{graphicx}
\usepackage{amssymb}
\usepackage{subfigure}
\usepackage{epstopdf}
\usepackage{graphics,psfrag}
\usepackage{graphicx,psfrag}
\usepackage{dcolumn}
\usepackage{bm}

\makeatletter
\begin{document}

\title{Evidence of a new quantum state of nano-confined water  }

\author{G. F. Reiter}
\affiliation{Physics Department,
  University of Houston,
  Houston, TX 77204, USA}
  \author{A.I. Kolesnikov}
\affiliation{Neutron Scattering Sciences Division, Oak Ridge National Laboratory, Oak Ridge, TN 37831, USA}
\author{S. J. Paddison}
\affiliation{Department of Biomolecular Engineering, University of Tennessee, Knoxville, TN, USA}
\author{P. M. Platzman}
\affiliation{80 Addison Road, Short Hills, NJ, 07078, USA}
\author{A.P. Moravsky}
\affiliation{MER Corporation, 7960, South Kolb Road, Tucson, AZ 85706, USA }
\author{M. A.  Adams}
\affiliation{ISIS Facility, Rutherford Appleton Laboratory, Chilton, Didcot, Oxfordshire, OX11 0QX, United Kingdom}
\author{J. Mayers}
\affiliation{ISIS Facility, Rutherford Appleton Laboratory, Chilton, Didcot, Oxfordshire, OX11 0QX, United Kingdom}

\date{\today}
\begin{abstract}

\noindent Deep Inelastic Neutron Scattering provides a means of directly and accurately measuring the momentum distribution of protons in water, which  is determined primarily by the protons  ground state wavefunction. We find that in water confined on scales of ~20$\AA$, this  wave function responds  to the details of the confinement, corresponds to a  strongly anharmonic local potential, shows evidence in some cases of coherent delocalization in double wells, and involves changes in zero point kinetic energy of the protons from  -40 to +120 meV difference from that of bulk water at room temperature. This behavior appears to be a generic feature of nanoscale confinement. It is exhibited here in  16 $\AA$ inner diameter carbon nanotubes, two different hydrated proton exchange membranes(PEMs), Nafion 1120  and Dow 858, and has been seen earlier in xerogel and 14 $\AA$ diameter carbon nanotubes. The proton conductivity in the PEM samples correlates with the degree of coherent delocalization of the proton. 
\end{abstract}
\maketitle
The properties of water that make it particularly suited to its role in sustaining life are the properties of the hydrogen bond network. This network is usually treated  as a collection of molecules, interacting weakly electrostatically. The hydrogen bonds that form are also regarded as primarily electrostatic. This picture is sufficiently accurate to give an excellent description of the structure of water, as measured by the pair correlations of the ions, and has been successfully incorporated into  water models using empirical interaction potentials.\cite{Pantalei:2008p963} The quantum delocalization of the protons can be incorporated in these models as well, and gives only small quantitative changes in the pair correlations. Deep Inelastic Neutron Scattering(DINS), or Neutron Compton scattering measures the proton momentum distribution, determined primarily by the ground state of the proton in the combined intra and inter-molecular potential.  This is far more sensitive to the local potential felt by the proton than measurements of structure.  Measurements of the width of the momentum distribution in bulk water from room temperature to the supercritical state have shown that the momentum distribution  cannot be readily calculated from the electrostatic model, except in the supercritical phase, where the density is significantly lower than in bulk water, and the model might be expected to work best\cite{Pantalei:2008p963}.  Recent simulations using high order multipole expansions to represent the interactions of the water molecules and an essentially exact potential surface to represent the intramolecular potential, have demonstrated that this is not simply an inadequacy in the ability of the empirical potentials to capture correctly the electrostatic interactions\cite{Christian1}.  It is an inadequacy, revealed in the momentum distribution, of the electrostatic model itself. There must be significant overlap of the electronic wavefunctions of the donor and acceptor molecules in the hydrogen bond. Such an overlap has been seen in x-ray Compton scattering earlier\cite{isaacs:2011p1539,romero:2001p1537}.  The electronic structure of bulk water is then a connected network, and the possibility exists that the network can respond quite differently than the interacting molecule model would suggest. We present evidence here that this is occurring in nano-confined water, water confined on the scale of 20$\AA$. The departures of the momentum distribution of the protons from that of bulk water are so large,  that we believe that the nano-confined water can be properly described as being in a qualitatively different quantum ground state from that of bulk water.    We begin with water in 16 $\AA$ inner diameter double wall carbon nanotubes (DWNT), and then compare the results with water confined in PEM (Nafion and Dow858), and in previously measured xerogel, and 14 $\AA$ single wall carbon nanotubes (SWNT).

The measurement of the momentum distribution of protons and other light ions is made possible by having sufficient high energy neutrons (2-100 eV) available that the scattering  takes place in the impulse approximation limit\cite{Andreani:2005p966}. In this limit, the scattering occurs on such a short time scale (10-100x10$^{-18}$sec) that the forces from the surroundings on the proton do not have time to act  and the proton is effectively a free particle for the duration of the collision\cite{Hohenberg:1966p1304,Reiter:1985p1003}. The scattering function then simplifies to 
\begin{equation}
S(\vec{q},\omega)=\int n(\vec{p})\delta(\omega-\frac{\hbar q^2}{2M}-\vec{q}\cdot\frac{\vec{p}}{ M})d\vec{p}=\frac{M}{q}J(y, \hat{q})
\end{equation}
where n($\vec{p}$) is the momentum distribution of the protons, $\hbar\omega$ and $\hbar\vec{q}$ the transferred energy and momentum, respectively, and M the mass of the proton.  The scattering function, which is directly proportional to the double differential scattering cross-section,  is  a function only of the rescaled variable $y=\frac{M}{q}\left(  \omega - { \frac{\hbar q^2}{2M}} \right)$,  the longitudinal momentum, and the direction of the vector $\vec q$. In our case, where all the samples are isotropic, it depends only on y and we have $S(\vec{q},\omega)=\frac{M}{q}J(y)$. No information is lost in J(y), which may be inverted to obtain n(p)\cite{Reiter:1985p1003,Andreani:2005p966}.    The instrument used was Vesuvio, at ISIS, the pulsed neutron source at the Rutherford-Appleton Laboratory in England. In the current experiment, we use 48 detectors, arranged with scattering angles from 35 to 75 deg on both sides of the beam. The data are corrected for multiple scattering\cite{mayers:2002p454} and a gamma ray background\cite{mayers:2011p015903}. The parameters of the fit are obtained by simultaneously fitting the data from all the detectors. 

The DWNT material was synthesized by chemical vapor deposition technique. The raw nanotube SWNT sample was produced by direct current arc vaporization of graphite-metal composite anodes. The metal component consisted of Co/Ni catalyst in a 3:1 mixture. The subsequent purification with hydrochloric acid was followed by the oxidation of non-tube carbon components in air at 300-600$^\circ$C. These preparation and purification steps produced micrometer-long nanotubes of a high purity, that is, with low metal catalyst content and low non-tube carbon content. The nanotube ends were opened by exposing the purified material to air at 420$^\circ$C for about 30 min. The samples were characterized by transmission electron microscopy (TEM) and small-angle neutron diffraction. For both the SWNT and DWNT, the (0,1) reflection of the two-dimensional hexagonal lattice of the bundle was evident in the diffraction data. The mean diameter of the SWNT was 14$\pm$1$\AA$, and the mean inner and outer diameters of the DWNT were 16$\pm$3$\AA$ and 23$\pm$3$\AA$ , respectively. The water absorption in SWNT and DWNT samples was controlled as follows: a mixture of de-ionized water and the nanotube material was equilibrated for 2 h in an enclosed volume at 110$^\circ$C; excess water was then evaporated at 45$^\circ$C until reaching the targeted water mass fraction. In this work, 3.0 g of SWNT and 2.8 g of DWNT samples were loaded with 11 and 13 wt.$\%$ of water, respectively.

We show in Fig.[1] the measured J(y)  for water in DWNT (the present work) and SWNT (measured earlier \cite{Reiter:2006p1262}), both at a temperature of 170 K, and bulk water at 300K. 
\begin{figure}[htp]

	\includegraphics[width=.5\textwidth,height=.375\textwidth]{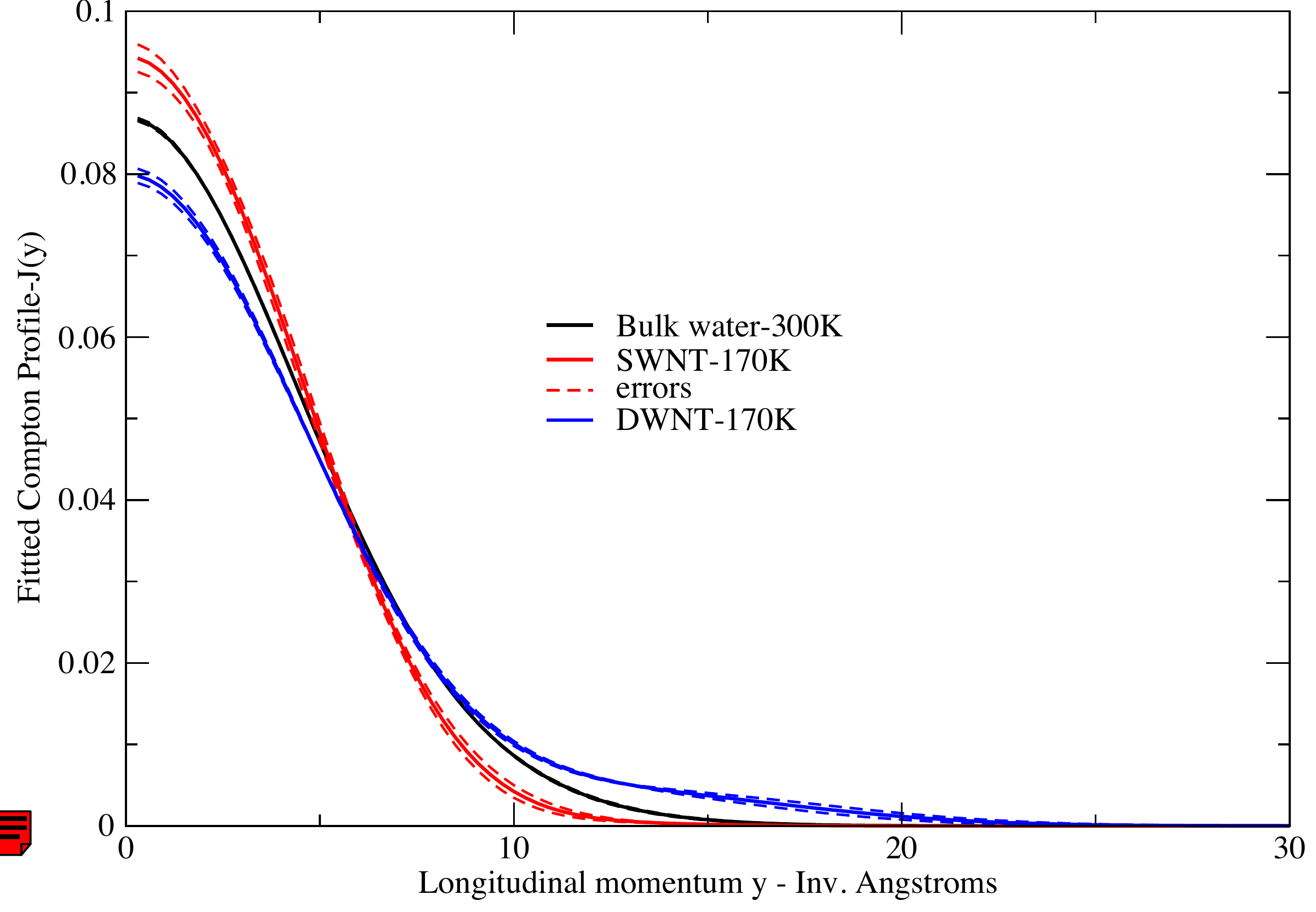}
	\label{fig:njy}

\caption{The measured Compton profile J(y) for water in 14$\AA$ SWNTs at 170 K, 16$\AA$ DWNTs at 170 K and bulk water at 300 K. The dashed lines are one standard deviation error bars.   }
\label{fig:jydist}
\end{figure}
  
Compared to bulk water,  the width of the Compton profile narrows in the case of water in the SWNT, and broadens for the water in the DWNT. These are essentially smooth cylindrical containers for the water, with diameters differing on average by only  two Angstroms, and similar average densities of water in the nanotubes , and yet the water responds completely differently in the two cases to the confinement.   
The structure of the water at 170 K in these systems, as determined by classical simulations using empirical potential models, consists of a cylindrical square ice sheath, about 3$\AA$ in from the carbon, with, in the case of the SWNT, a single chain of water molecules down the center,\cite{Kolesnikov:2004p1291}  and in the case of the DWNT, a smaller inner cylinder of water down the center\cite{Striolo:2005p1409}. In both cases, most of the water is in the ice sheath.  We conclude  that the quantum state of the protons in confined water is very sensitive to the nature  of the confinement, and responds to the global configuration of the hydrogen bond network. 
 
  Further differences between the two systems, and further evidence of the sensitivity of the hydrogen bond network to confinement,  are evident by comparing the temperature dependence of the proton momentum distributions. As reported in an earlier work\cite{Reiter:2006p1262}, the momentum distribution in the SWNT was temperature independent, to within the errors of our measurement, up to 230 K. Fig.[1] would not look observably different if we replaced the SWNT data at 170 K, with that at 5 K. The DWNT data are, however, strongly temperature dependent. We show in Fig[2] the momentum distributions as a function of temperature in the DWNT. DINS is sensitive to coherent delocalization, the spreading out of the wave function into a bimodal distribution as in a double well system.   The oscillation seen in the momentum distribution is characteristic of such a system\cite{Reiter:1985p1003,Andreani:2005p966}. Evidently, a significant number of the protons in DWNT are in that state, and further, the separation of the wells, which can be inferred from the position of the minimum (~$\pi$/$p_{min}$), is changing with temperature, getting smaller from 4.2K to 120K, and then getting larger at 170K. 
  
  One might think that the dramatic changes in the proton momentum distribution were a peculiarity of the structure of the water in the nanotubes at low temperature. At room temperature, that structure disappears, and the water resembles locally the fourfold coordinated hydrogen bond network of the bulk. We see in Fig.[2] however that significant differences remain at room temperature between the bulk and nano-confined water.
   
   The interest in water in carbon nanotubes is partially as a simple model for biological channels. As such, the quantum state of the protons at room temperature is particularly relevant. As is evident from Fig[2] this state is far from that of bulk water.  We see from Table 1 that the kinetic energy/proton is higher in the confined space of the nanotube by 34meV than it is in bulk water, or roughly kT at room temperature. Since there are two protons in each water molecule, the changes of the proton zero point energy with confinement  is clearly significant for water entering and leaving the channels. The quantum state of the water in the channels may also be of significance for transport through the channels, which is orders of magnitude greater than continuum fluid flow theories would predict.\cite{Majumdar:2005p1535} 

It might be thought that the appearance of double well potentials for the protons in nano-confined water was a result only of the specific ordered structures at low temperature in the nanotubes. This is not the case. The phenomenon is more general.   Carbon nanotubes are smooth, and the interaction of the water with  the surface of the nanotube is primarily Coulomb repulsion and Van de Waals attraction. There is no chemical binding to speak of between the water and the nanotube walls. These conditions are not present in an earlier room temperature experiment that was done on the confinement of water in the pores of xerogel\cite{Garbuio:2007p939}, which can be thought of as a glass sponge. The surface of the pores contains Si-O-H (silanol)groups, to which the water can hydrogen bond, and the proton can be ionized and become mobile in the confined water. It was observed that the momentum distribution of the water in nominally 23$\AA$ pores could be described as though all the molecules in the pore were confined in a double well potential.  For larger pores, 80$\AA$, it appeared that the average momentum distribution was much closer to that of bulk water. The interpretation at the time was that the water in the surface layer was being affected by the interaction with the silanol groups and somehow the surface effects propagated throughout the water in the small pore. That is that the result was due to a particular interaction of the surface with the water. Given the results in the nanotubes, we think now that the effect on the water, rather than being due to local interactions at the surface with the silanol groups , was a direct effect of the confinement itself, and that 20$\AA$ is the scale size at which this effect becomes apparent.  

Two systems that are similar in some respects to the xerogel system are the perfluorosulfonic acid membranes: Nafion 1120 and Dow 858\cite{Kreuer:2008p1390}. These are ionomers with hydrophobic poly(tetrafluoroethylene) (PTFE) backbones and randomly pendant perfluoroether side chains terminating with sulfonic acids. The ionomers when hydrated exhibit a sponge like structure, a nanophase separated morphology where the water and ions exist in domains that are only a few nanometers in diameter surrounded by the backbones. The sulfonic acid group (-$SO_{3}H$) donates its proton to the water when there is sufficient water in the pores of the sponge, making them very good proton conductors.   We show in Fig.[3] the momentum distribution at room temperature for the water confined in the pores of the two membranes, at the same water content (14$H_{2}O/SO_{3}H$), presented here for the first time.  The deviation from the bulk water ground state in these two samples is dramatic and corresponds to a kinetic energy difference of 107meV/proton for Nafion and 124 meV/proton for Dow 858. They are not even qualitatively in the same state as bulk water.

\begin{figure}[htp]

	\includegraphics[width=.5\textwidth,height=.375\textwidth]{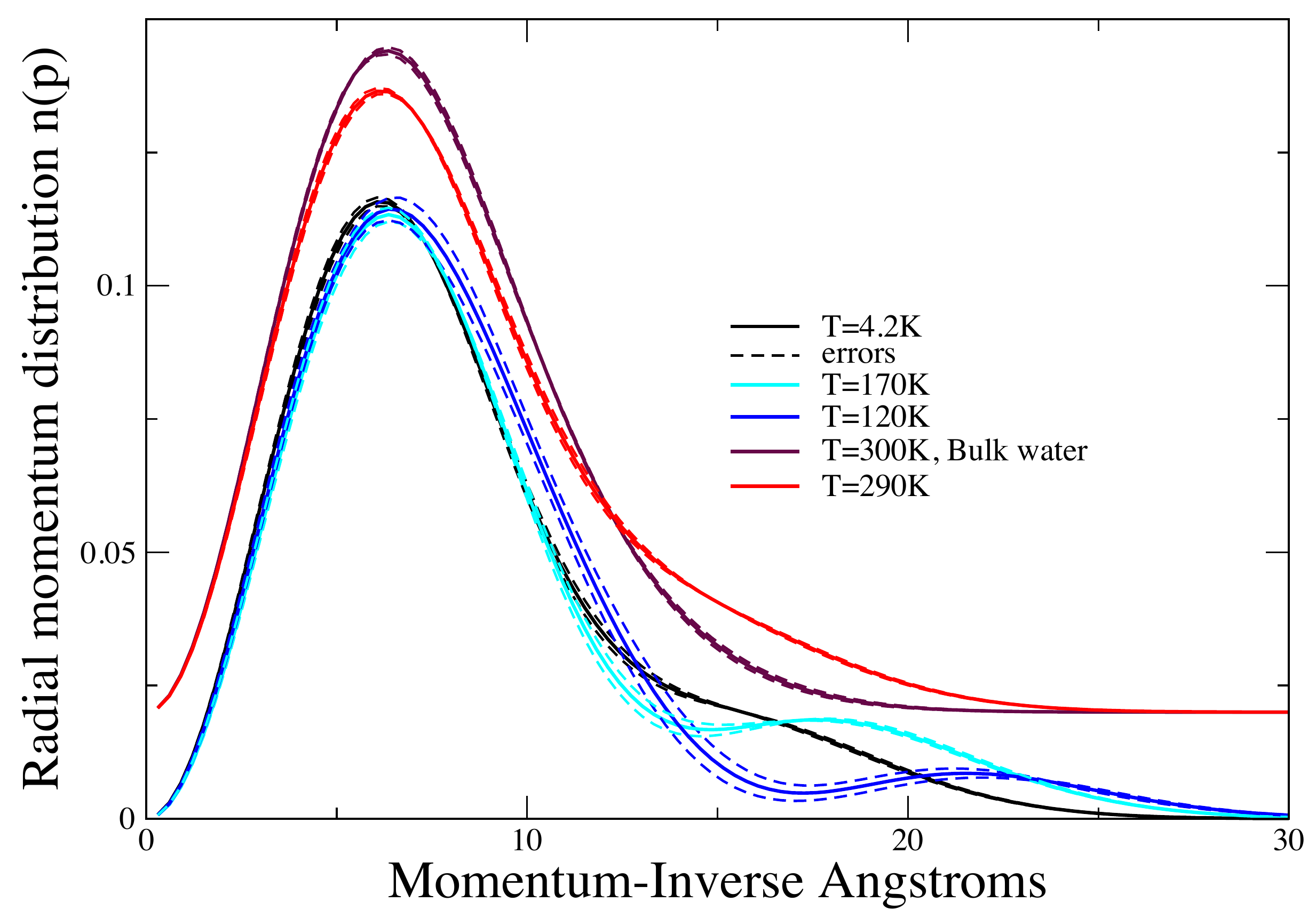}
	\label{fig:dwmom}

\caption{The momentum distribution of the water protons in 16 $\AA$ DWNT as a function of temperature, compared with that of bulk water at room temperatures. The nanotube 290K signal and the bulk water signal have been displaced upwards by .02 units for clarity. }
\label{fig:momdist}
\end{figure}

\begin{figure}[htp]

	\includegraphics[width=.5\textwidth,height=.375\textwidth]{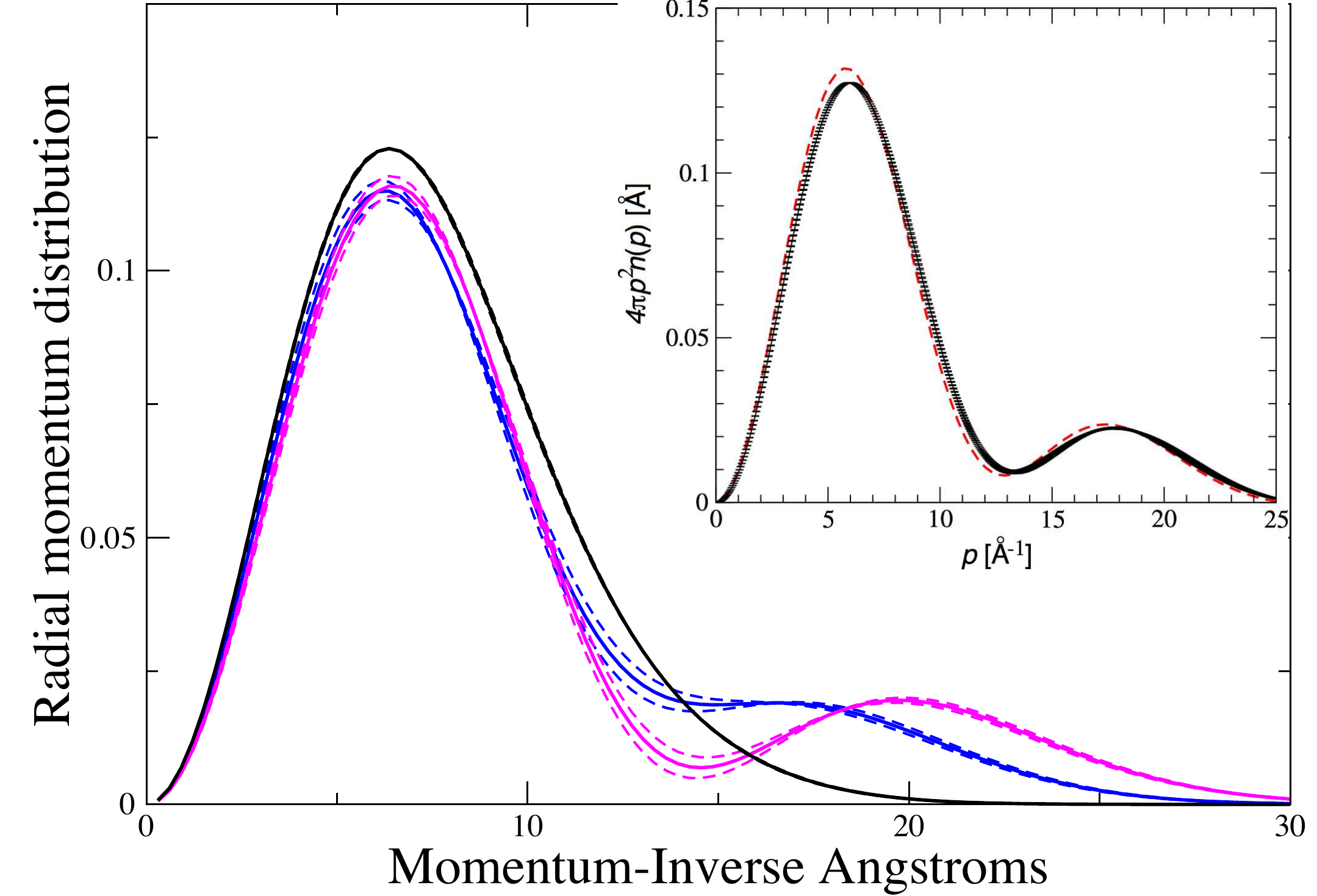}
	\label{fig:dwmom}

\caption{The momentum distribution of the protons in Nafion1120(blue) and Dow 858(magenta) compared with that of bulk water(black), all at room temperature. There are 14 water molecules/sulfonyl group in both materials. The inset is the momentum distribution of water in the 23$\AA$ pores of xerogel at room temperature. The dashed red line is a fit to the data with a single particle in a double well model.\cite{Garbuio:2007p939}}
\label{fig:momdist}
\end{figure} 
The water confined in the PEM's is highly acidic, and it might be thought that the acidity is somehow primarily responsible for the dramatic difference between the bulk and confined water. We have observed DINS for bulk HCl solutions at these concentrations, and found only small deviations from that of bulk water\cite{icns}. We think there is no doubt that it is not the free protons that are driving the transition from one quantum state to the other, but the confinement itself. The interaction with the surface groups, the sulfonyl and silanol groups, may, however  play a role in the large difference between the nanotube data and that in Nafion and xerogel.\cite{Habenicht:2010p1548, Habenicht:2010p1547} 

The proton conduction mechanism  in PEMs is thought to be at least partially due to the Grothuus mechanism\cite{Kreuer:2004p355}. This is essentially a classical mechanism.  The protons in the confined water are largely in a state that appears to be bimodally, coherently, distributed in space, and are therefore far from behaving as classical particles. It is  likely that the proton transport in these materials is significantly  different from the process in bulk water. It is intriguing that the conductivity of the Dow 858 is higher than that of the Nafion by 70$\%$\cite{Kreuer:2008p1390}, which correlates qualitatively with the proportion of protons involved in  coherent delocalization seen in Fig.[3], indicating that coherent transport through a series of delocalized proton states may be an efficient conduction mechanism. 

In our earlier experiments on SWNT and xerogel\cite{Reiter:2006p1262, Garbuio:2007p939} we have analyzed the data in terms of a single particle in a potential model. This model can provide a good phenomenological fit, as can be seen from the inset in fig 3. We believe, however,  that this  model is inadequate. Vibrational densities of states measurements in both sets of nanotubes have shown a slight blue shift of the proton-oxygen  stretch mode.\cite{Kolesnikov:2004p1291}  It is not possible to explain this within a single particle model. The softening of the potential seen in the momentum distribution, although  consistent with the increase in the Debye Waller factor that was observed earlier \cite{Kolesnikov:2004p1291}, is too extreme to accommodate an increase in the ground state to first excited state stretching  transition frequency.   We believe that the sensitivity to confinement, and the large departures from the momentum distribution expected for a single water molecule are due to correlated proton motion in the ground state of the system, and expect that there is a quantum coherence length associated with these correlations. There is experimental confirmation of correlated proton motion in the vibrational spectrum of ice \cite{Shultz:2010p1533, Shultz:2010p1532} as well as  earlier predictions of this effect  from simulations
\cite{Buch:2005p1420}. 

Whatever the origin of the effects we have observed, it is clear that they can be present in room temperature confined water, that they respond to the details of the structure of the hydrogen bond network of the water, they indicate a strongly anharmonic state, and that the zero point energy shifts can be  comparable or greater than kT. The changes in the zero point motion of the protons in confined water, as in living cells for instance, can be expected to play a significant role in the energetics of the cells, where typical distances between components are on the order of 20$\AA$. Changes in zero point energy have already been shown to play a major role in the binding of water molecules to DNA\cite{Reiter:2010p1546} The change in the environment for proton transport in the PEM systems such as Nafion can be expected to have a significant effect on the conductivity of these materials.  

\begin{table}
\small

\caption{Parameters for fits to the momentum distributions. 
n(p)= $\frac{1}{(\sqrt{2\pi}\sigma)^3} e^{-\frac{p^{2}}{2\sigma^{2}}}
\cdot\Big[1+\sum_{n=2}^{\infty}a_n (-1)^nL^{\frac{1}{2}}_{n}(\frac{p^2}{2\sigma^2}) \Big]$\\ where $L^{\frac{1}{2}}_{n}(\frac{p^2}{2\sigma^2})$ is an associated Laguerre poynomial. p is in units of $\AA^{-1}$.
K. E. is the kinetic energy, $\frac{3\hbar^2}{M}\sigma^2$. 
 } 
 
  \begin{tabular}{|c|c|c|c|c|cl}
\hline
Water Sample& $\sigma (\AA^{-1})$ &K.E.-meV&$a_2$& $a_3$&$a4$\\
\hline
Bulk-300K&$4.72\pm.03$&138&$.258\pm .060$&0.0&0.0\\
\hline
 DWNT-4,2K&$5.40\pm.36$&181&$1.20\pm.58$&0.0&0.0\\
\hline
DWNT-120K&$5.65\pm.06$&198&$2.44\pm.31$&$2.05\pm.15$&$1.07\pm.16$\\
\hline
DWNT-170K &$5.97\pm.15$&221&$1.84\pm .19$&$0.0$&0.0\\
\hline
DWNT-290K &$5.26\pm.08$&172&$0.98\pm .13$&$0.0$&0.0\\
\hline
SWNT-170K & $4.09\pm.07 $&104&0.0& $.35\pm.22 $&$0.0$\\
\hline
Nafion1120-300K & $6.28\pm.40$ &245&$2.42\pm.51 $& 0.0&$0.0$\\
\hline
Dow858-300K & $6.50\pm.27$ &262&$2.70\pm.36 $&0.0&$0.0$\\
\hline

\end{tabular}
\end{table}


\centerline{Acknowledgements}
G. Reiter's  work was supported by the DOE, Office of Basic
Energy Sciences under Contract  No.DE-FG02-08ER46486. Work at ORNL was managed UT-Battelle, LLC, for the DOE under contract DE-AC05-00OR22725. The STFC Rutherford Appleton Laboratory is thanked for access to neutron beam facilities.
\bibliography{bib8}

\end{document}